# Asymmetric Electrostatic Dodecapole: Compact Bandpass Filter with Low Aberrations for Momentum Microscopy

O. Tkach[1,2], S. Chernov[3], S. Babenkov[1], Y. Lytvynenko[1,4], O. Fedchenko[1], K. Medjanik[1], D. Vasilyev[1], A. Gloskovskii[3], C. Schlueter[3], H.-J. Elmers[1] and G. Schönhense[1]*

*1 Johannes Gutenberg-Universität, Institut für Physik, D-55099 Mainz, Germany*
*2 Sumy State University, Rymskogo-Korsakova 2, 40007 Sumy, Ukraine*
*3 Deutsches Elektronen-Synchrotron DESY, D-22607 Hamburg, Germany*
*4 Institute of Magnetism of the NAS and MES of Ukraine, 03142 Kyiv, Ukraine*
** corresponding author: schoenhe@uni-mainz.de*

## Abstract

Imaging energy filters in photoelectron microscopes and momentum microscopes use spherical fields with deflection angles of 90°, 180°, and even 2 x 180°. These instruments are optimized for high energy resolution, but exhibit image aberrations when operated in high transmission mode at medium energy resolution. Here we present a new approach for bandpass-filtered imaging in real or reciprocal space using an electrostatic dodecapole with an asymmetric electrode array. In addition to energy-dispersive beam deflection, this multipole allows aberration correction up to the 3rd order. Here we describe its use as a bandpass prefilter in a time-of-flight momentum microscope at the hard X-ray beamline P22 of PETRA III. The entire instrument is housed in a straight vacuum tube because the deflection angle is only 4° and the beam displacement in the filter is only ~8 mm. The multipole is framed by transfer lenses in the entrance and exit branches. Two sets of 16 different sized entrance and exit apertures on piezomotor driven mounts allow selection of the desired bandpass. For pass energies between 100 and 1400 eV and slit widths between 0.5 and 4 mm the transmitted kinetic energy intervals are between 10 eV and a few hundred eV (FWHM). The filter eliminates all higher or lower energy signals outside the selected bandpass, significantly improving the signal-to-background ratio in the ToF analyzer.

## 1. Introduction

Imaging bandpass filters play an important role in photoelectron spectroscopy. Such filters can either transport either real-space images, as studied in detail by Tonner [1], or momentum-space images, as pioneered by Kirschner and coworkers [2]. The latter family of instruments established a new type of angle-resolved photoelectron spectroscopy (ARPES), called momentum microscopy (MM). This early work initiated the development of several types of momentum microscopes, whose front lens converts the photoelectron angular distribution into a full-field image of the transversal ($k_x$, $k_y$) momentum pattern. Dispersive-type MMs use either a tandem arrangement of two hemispherical analyzers [2-6] or a single hemispherical analyzer [7-10]. These instruments are similar to the spectroscopic low-energy electron microscope (SPELEEM) as developed by Bauer et al. [11, 12]. An alternative design using the dispersive power of a magnetic prism has been developed by Tromp et al. [13].

In all of these instruments energy-filtered 2D images (real or reciprocal space) are transported through the analyser using concepts from charged particle optics and electron microscopy. As in conventional spectrometers, the energy resolution is defined by the pass energy and the size of the entrance and exit slits. A resolution of 4.2 meV has been achieved using a single hemisphere setup [8]. Another type of momentum microscope is based on time-of-flight (ToF) recording [14]. ToF-MMs simultaneously acquire an energy band of several eV width using a time-resolving image detector, allowing 3D ($E_{kin}$, $k_x$, $k_y$) recording.

The hemisphere-based instruments have one or two 180° deflections [1-10] or one 90° deflection [11, 12] with corresponding transfer lens systems. Thus, these instruments are associated with image aberrations when they are operated in high transmission mode with medium energy resolution. Here, we present a new design of a compact electrostatic bandpass filter that fits into the linear column of a ToF-MM. We adopt the concept of an electrostatic dodecapole, as theoretically studied in detail by Boerboom et al. [15, 16]. This type of multipole allows the superposition of dipole, quadrupole and hexapole fields and thus can correct image aberrations up to the 3$^{rd}$ order.

A prototype dodecapole is used as a prefilter in a ToF-MM at the hard X-ray beamline P22 at PETRA III, where it significantly reduces the background signal and eliminates artifacts in the ToF spectra. At high kinetic energies, the so-called 'temporal aliasing' effect [17] can lead to artifacts in spectra and momentum patterns. These artifacts are caused by electrons, whose time of flight differs from the flight time of the probe electrons of interest by a multiple of the pulse period of the photon source. The time-resolving detector cannot distinguish such (faster or slower) background electrons from the true signal. The energy dispersive bandpass filter limits the recorded energy interval to the region of interest, thus completely eliminating the aliasing artifacts.

The dodecapole arrangement consists of an entrance and an exit branch, both of which contain sets of 16 size-selectable and position-adjustable apertures that serve as entrance and exit apertures. The transmitted energy band depends on the pass energy, slit width and deflection angle. When used for pre-filtering, the dodecapole unit must not be operated with high energy resolution, as the desired final resolution is reached by the ToF analysis. For the prototype at beamline P22, a small deflection angle of only 4° was chosen. This allows the use of a linear vacuum housing with mumetal tube for magnetic shielding.

## 2. The asymmetric dodecapole: A versatile device for energy filtering and aberration correction

### 2.1 Motivation for the development of the dodecapole filter

The motivation for the development of the dodecapole with asymmetrical electrode arrangement originated from an inherent problem of the ToF method. ToF analyzers act as high-pass filters: All electrons with energies above a certain cutoff, defined by the potential of the drift tube, can pass the ToF section and reach the detector. The time-resolving detector is triggered by the bunch clock of the pulsed source and the trigger pulses come in with the period $T$ of the photon pulses. Hence the detector cannot distinguish electrons with time-of-flight $\tau$, $\tau$+n$T$ or $\tau$-n$T$ (n, natural number). The period $T$ is thus the crucial quantity. The majority of storage rings use filling patterns with $T$=2ns (500MHz); a few, e.g. MAX IV, at 10ns (100MHz). Special filling patterns are provided by PETRA-III in 40-bunch mode with $T$=183ns (~5MHz) or BESSY-II in single-bunch mode with 1μs (1MHz). The FELs FLASH and European XFEL, both hosting ToF MMs, provide micro-bunch trains with 1MHz. Laboratory sources range between 16ns (60MHz) for intra-cavity enhanced high-harmonic sources [18] and the ~10μs range (~100 kHz) (e.g. [19]). The time-of-flight $\tau$ of the electrons in the ToF analyzer can exceed the period $T$. Fast electrons released by the pulse at $t_0$ can overtake slow electrons from earlier pulses at $t_0$-n$T$. This 'temporal aliasing' effect is quantified for realistic conditions in the hard-X-ray range in Ref. [17], Fig. 4(f).

To understand how the dodecapole eliminates this effect, we first need to examine the functioning principle of a ToF-MM. Figures 1(a, b) and (c, d) display the ToF Photoemission Electron Microscopy (ToF-PEEM) mode and the ToF-MM mode, respectively. Figs. 1(a, b) show the two fundamental rays: the axial ray $u_\alpha$, which starts on the axis at an angle $\alpha$, and the field ray $u_\pi$, which runs parallel at a distance from the optical axis. To enhance visibility, we display ray bundles around $u_\alpha$ in black and around $u_\pi$ in blue. Additionally, a green field ray is shown at half the distance from the axis. The first k-image is formed in the backfocal plane (also known as the reciprocal plane RP1) of the front lens, as detailed in Fig. 1(c). The front lens and lens 1 create a real-space image of the sample in the first Gaussian plane GP1. The $u_\alpha$ ray intersects with the axis, while the field rays (blue and green) run parallel to the optical axis, determining the lateral magnification. The zoom lens 2 can adjust the magnification between Gaussian planes GP1 and GP2, and lens 3 produces the final PEEM image in GP3 (at the delay-line detector DLD). To enhance resolution and contrast, a contrast aperture CA can be inserted in plane RP1 (top left of (b)). The low-energy drift section, necessary for energy dispersion, ensures a large final magnification. The simulation provided is for a realistic lens system with a 200 μm field-of-view (blue rays in (a)) and magnifications of $M_1$=11, $M_2$=42, and $M_3$=330 in GP1, GP2, and GP3, respectively.

In the $k$-imaging mode, Fig. 1(c, d), the region on the sample is typically an order of magnitude smaller, either due to a small photon spot or due to the confinement of the analysed region by a field aperture FA in GP1, see Fig. 1(d). However, the observed $k$-range is much larger than in PEEM mode. The zoom lens 2 and transfer lens 3 are adjusted so that the final reciprocal plane RP3 is shifted towards the detector. At a given time $t$, there is a certain distribution of electrons on their way to the detector. The coloured contours of the isochrones mark three different types of electrons: 1. The signal of interest (released by the photon pulse at $t_0$); 2.

The slow electron background (from the previous pulse $t_0$-$T$) and 3. Fast core-level photoelectrons from the next pulse ($t_0$+$T$) or from core levels with smaller binding energy. Faster electrons often originate from higher order admixtures in the photon beam. These fast core-level electrons are underfocused, while the slow background electrons are overfocused, both of which result in a smaller diameter of the *k*-pattern.

Figures 1(e, f) show an experimental example measured at beamline P22 at PETRA III (DESY, Hamburg) at h$\nu$ = 4 keV. It shows the expected behaviour. The different species 1, 2 and 3 appear with different diameters in the $k_x$-$k_y$ pattern (e). The $\tau$-vs-$k_{\|}$ section (f) shows the low-energy electrons with their typical parabolic rim and the fast core-level electrons as a band with parallel rims. A series of core levels appear as horizontal stripes. Figure 1(g) shows a scheme of the temporal structure for this case. The blue, green and red signals fall in the same time interval, although they originate from different photon pulses and have different energies. We note that this case with two parabolic rims is rather an exception.

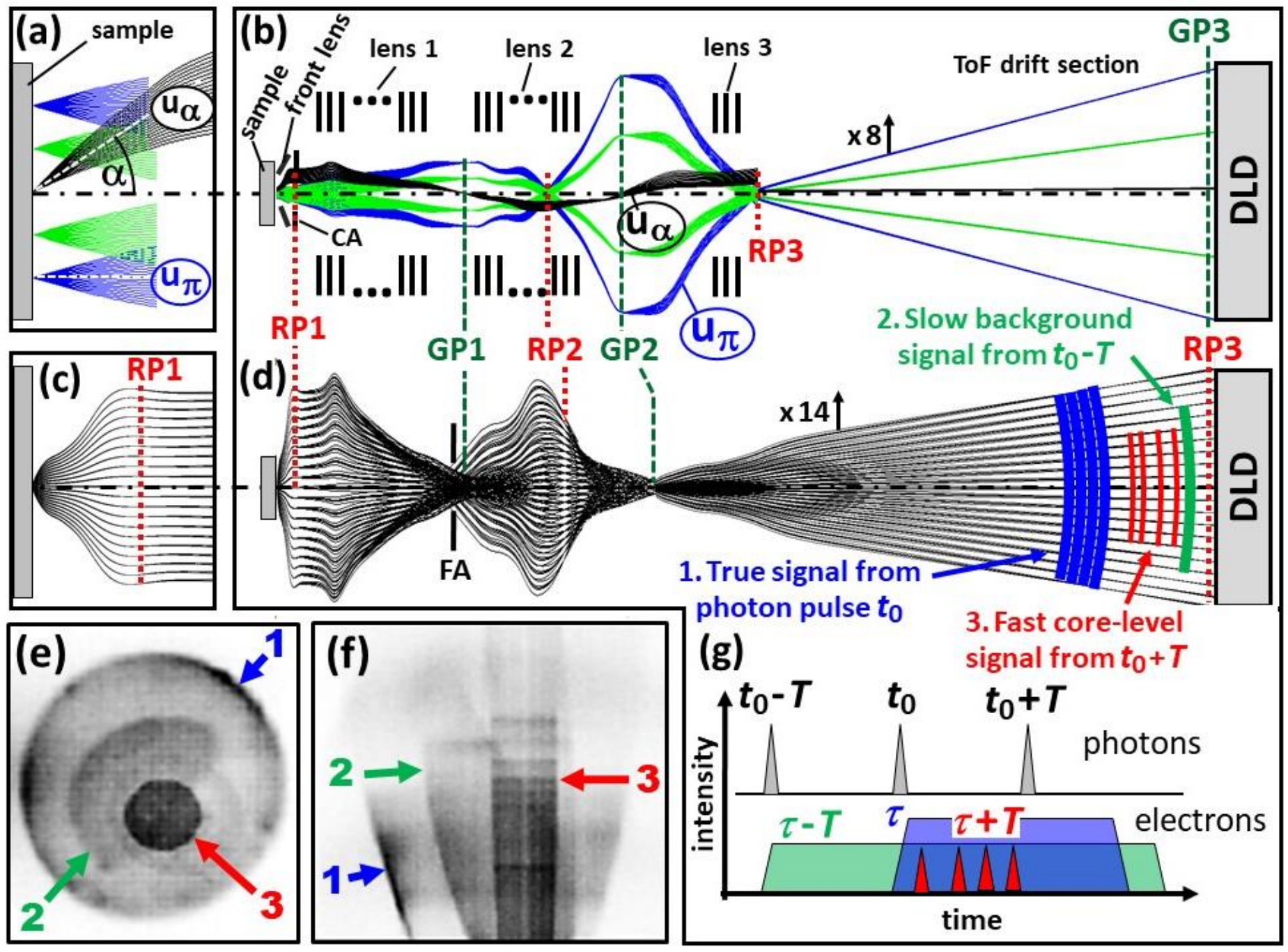

**Figure 1**

Working principle of a time-of-flight photoelectron microscope, demonstrated by ray tracing calculations for a typical microscope geometry. (a, b) Photoemission electron microscopy (ToF-PEEM) mode, illustrated by ray bundles around the fundamental rays $u_\alpha$, starting on the axis at an angle $\alpha$, and $u_\pi$, starting parallel to the optical axis, as seen in the close-up view (a); angles are greatly exaggerated. Lenses 1, 2 and 3 are multi-element groups as schematically sketched; stigmators and deflectors are omitted. RP1, RP2 and RP3 (red dotted lines) are the reciprocal image planes; GP1, GP2 and GP3 (green dashed lines) are the Gaussian (real space) image planes; DLD is the delay line detector. (c, d) Same as (a, b) but for the momentum microscopy mode. The first *k*-image appears in the plane RP1 (backfocal plane of the front lens), marked in (c). In (b) and (d), the ToF drift sections are radially compressed by factors of 8 and 14, respectively. (e) $k_x$–$k_y$ and (f) $\tau$-*vs*-$k_{\|}$ sections of a measured momentum pattern with three overlapping signals due to the 'temporal aliasing' effect. The largest pattern (1) corresponds to the electrons of interest with time-of-flight $\tau$. Patterns (2) and (3) result from slower and faster electrons with time-of-flight $\tau$+$T$ and $\tau$-$T$, where $T$ is the period of the photon pulses. A series of fast core-level photoelectrons is visible as horizontal stripes in signal 3. (Ray tracing using SIMION 8.0 [20]).

In summary, faster and slower electrons can be detected in the same time interval as the true signal, if their time-of-flight differs from that of the true electrons by a multiple of *T*. This can lead to significant artifacts in the momentum distributions because the time-resolving detector is unable to distinguish these false 'modulo *T*' electrons from the true signal. The only way to eliminate these background signals is to implement a dispersive energy filter. It has been demonstrated that a hemispherical analyser can accomplish this task [8]. However, the electrons traveling in the spherical field experience significant time lags that require correction. The goal of this study was to create a compact dispersive bandpass filter with minimal aberrations that does not affect the time structure of the electron signals.

## 2.2 The asymmetric dodecapole as versatile electron-optical element

We have chosen the dodecapole as dispersive prefilter because it can generate a number of multipole fields for aberration correction up to 3rd order. The particle-optical properties of twelve electrodes in a dodecagon configuration have been studied in detail by Boerboom et al. [15, 16]. The most important features are: deflection in two perpendicular directions by dipolar fields, quadrupole focusing and stigmatization, correction of second- and third-order aberrations, and two-directional focusing. A unique advantage is that several of these applications can be effected simultaneously by superposition of multipole field arrays. First-, second- and third-order focusing can be independently controlled by appropriate settings of the voltages applied to the electrodes. The influence of the fringing field is of fourth-order and thus disregarded in the third-order approximation.

Figure 2(a) shows a schematic view of the electrode arrangement. Depending on the size of the recorded *k*-field-of-view and the pass energy, the beam diameter in the momentum image located in the centre of the dodecapole can be very large. Therefore, its geometry has been chosen such that the deflecting field is homogeneous in a range as large as possible. Otherwise the *k*-image would be distorted upon deflection. Figure 2(b) shows the asymmetrical arrangement and voltages of the electrodes with the centre as zero reference. The electrode geometry has been designed such that for a pure deflection field the voltages vary linearly: U, 2/3 U, 1/3 U, 0, -1/3 U, -2/3 U, -U. The calculated equipotential contours reveal that the field along the dispersive direction (vertical) is homogeneous in a large area of ~ 80% of the total cross section (marked by the dashed circle). For comparison, Fig. 2(c) shows the electric field in a conventional octupole unit with the same outer dimensions as in (b), assuming that the centres of the rods lie on the outer contour of the unit. Here the homogeneous region is much smaller, ~ 50% of the total cross section.

Figure 2 shows the potential distributions of the two perpendicular dipolar fields (b, d) and the two quadrupole fields (e, f), rotated by 45° with respect to each other. These have the function of (weak) cylinder lenses with selectable azimuthal orientation, compensating for astigmatism. Figs. 2(g, h) show equipotential lines for the superposition of the primary dipole field (b) with the differently oriented quadrupole fields (e, f).

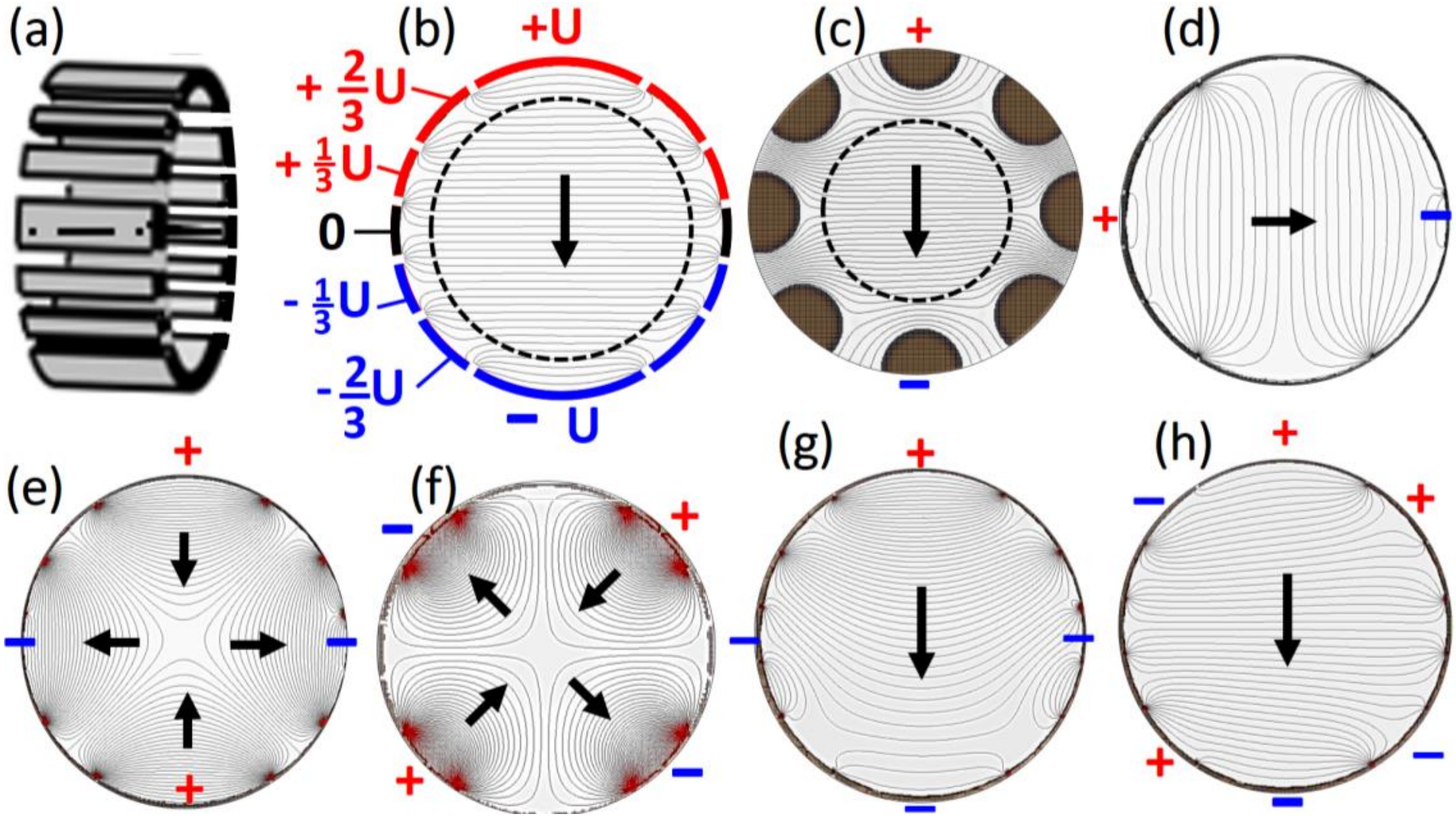

**Figure 2**

(a) Schematic view of the dodecapole bandpass filter. (b) Electrode geometry, voltages and calculated potential distribution in the center cross-sectional plane of the dodecapole; the dispersive plane is vertical. The arrow denotes the deflecting (dipole) field, which causes the energy dispersion. (c) Dipole field created by a conventional octupole deflector/stigmator with same outer dimension as (b), shown for comparison. (d-h) Correcting multipole fields: Horizontal dipole field (d) correcting for beam position; quadrupole fields oriented at 0°/90° (e) and +/- 45° (f), correcting for astigmatism. (g) and (h), examples for the superposition of the deflector field (b) and the quadrupole fields (e) and (f), respectively.

## 2.3 Properties of the dodecapole bandpass filter

In the limit of small deflection angles, the homogeneous field (marked by the arrow in Fig. 2(b)) deflects the electrons along parabolic trajectories. We have performed ray tracing calculations in order to elucidate the resolving power for different geometries and deflection angles. Systematic simulations showed that an impact angle of 2° with respect to the symmetry axis of the dodecapole is a good compromise between sufficient resolution (when used as a bandpass *prefilter*) and small aberrations. For such a small angle, the fringing field effects at the entrance and exit of the dodecapole are negligible.

Figure 3 displays a selection of results for the $\alpha$ = 4° geometry of the instrument installed at beamline P22. The assignment of lenses and planes is consistent with Fig. 1. The dodecapole is placed between the Gaussian planes GP1 and GP2. The dodecapole unit with entrance and exit slits and transfer lenses, shown in Fig. 3(a), replaces lens group 2. The field aperture serves as *entrance aperture* and an identical array of 16 apertures in GP2 serves as *exit aperture.* This entrance aperture is equivalent to FA in Fig. 1 and has two additional functions: It selects the diameter of the observed region of interest on the sample, independent of the size of the illuminating photon beam. Second, in a momentum microscope FA has the function of a contrast aperture for the *k*-image. The transfer lens 2A produces a *k*-image with parallel rays in the center of the dodecapole (reciprocal plane RP2). Lens group 2B focuses an image from plane GP1 into the plane of the exit aperture GP2. The dodecapole is tilted by 2° and the exit branch is tilted by 4°. The red equipotential contours (equally spaced between the upper and lower electrodes) show that the homogeneous field in this section has significant distortions at the entrance and exit of the dodecapole. Two additional electrodes F1 and F2 ensure that the field penetration of the lenses 2A and 2B is symmetrical at the entrance and exit. The focal spot in plane GP2 shows no significant distortions.

Figure 3(b) shows seven groups of rays with different kinetic energies (inside of the dodecapole). Electrons with a kinetic energy of $E_{\mathrm{kin}}$ = 580 eV (= pass energy $E_{\mathrm{pass}}$) are focused to pass through the exit aperture (black rays). Three faster groups at 620, 660 and 700 eV (blue) and three slower groups at $E_{\mathrm{kin}}$ = 540, 500 and 460 eV (red) are deflected too weakly and too strongly, respectively. At $\Delta$E = 40 eV (7 % of $E_{\mathrm{pass}}$), all groups are completely separated. The dispersion in the exit plane is 5 $\mu$m/eV. Fig. 3(c) shows a simulation for a *k*-field diameter of 10 Å$^{-1}$ with a stretched radial coordinate, giving the appearance of an increased angle. Insert (d) shows the detail marked as a dashed rectangle in (c) after dewarping, i.e. with a true 4° deflection angle. The pattern shows that the field curvature is increased in the *k*-image in plane RP3 behind the exit aperture. However, there is no visible deformation of the *k*-image due to the dodecapole in the dispersive plane. The dodecapole field acts as a weak cylinder lens, compressing the pattern in the dispersive plane by about 7% compared to the non-dispersive plane, which can be corrected by image processing.

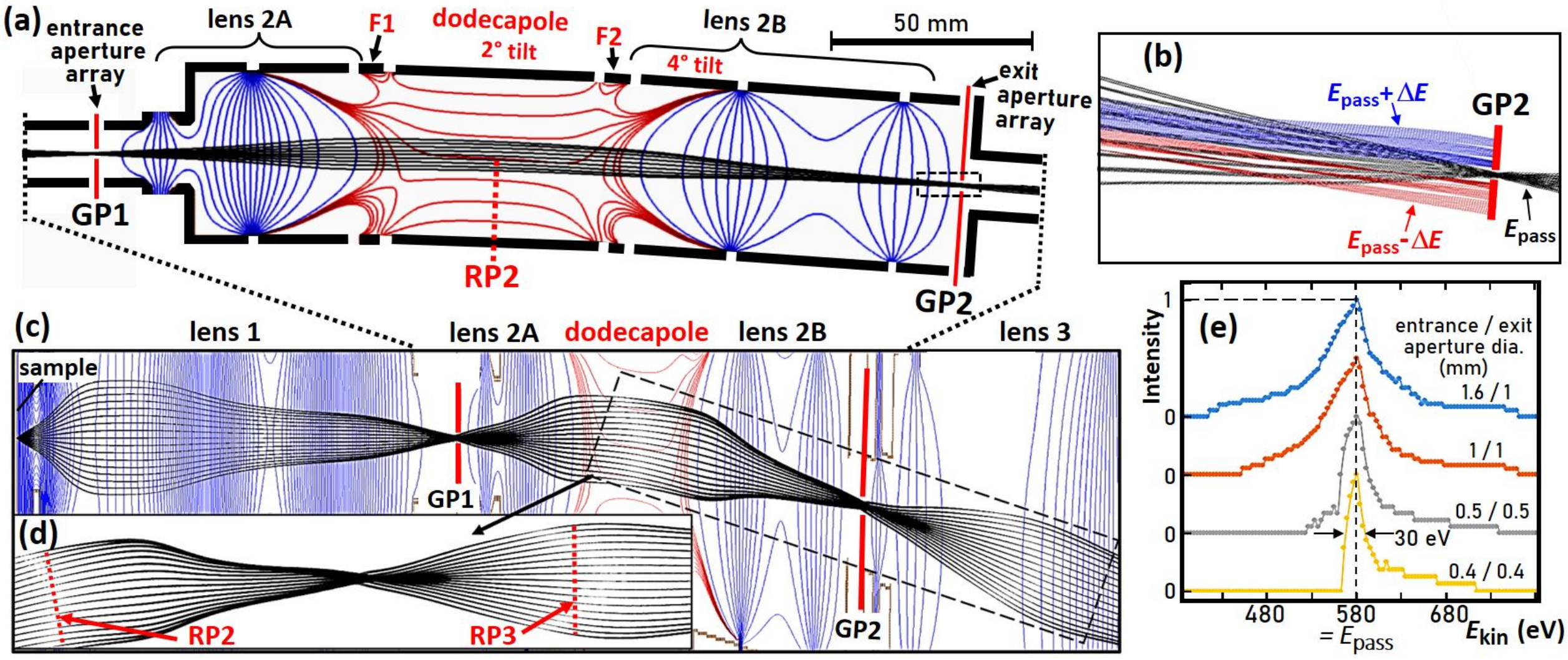

**Figure 3**

Ray tracing simulations of a full 3D model of the dodecapole bandpass filter, integrated into the lens system of the ToF momentum microscope at beamline P22 of PETRA III. (a) Scaled view (scale bar 50 mm) of the bandpass filter, consisting of entrance aperture (in plane GP1), lens 2A, fringe field correctors F1 and F2 framing the dodecapole, lens 2B and exit aperture (plane GP2). Equipotential contours for the lens fields in blue and for the dodecapole in red. (b) Close-up view of the region near the exit aperture for five different energies: $E_{\mathrm{kin}}$ = $E_{\mathrm{pass}}$ = 580 eV (black rays) and faster (blue rays) and slower electrons (red rays) with $\Delta$E varied in steps of 40 eV. (c) View including lens 1 and lens 3 with the radial coordinate stretched. The dashed rectangle marks the region shown in (d) after dewarping (conformal scale). (e) Series of calculated transmission profiles; see text for details. (Ray tracing using SIMION 8.0 [20])

Figure 3(e) shows a series of simulations of transmission profiles for various aperture sizes [FA / SA = 1.6 / 1.0, 1.0 / 1.0, 0.5 / 0.5, and 0.4 / 0.4; all in mm]. The profiles were calculated by running 5000 rays for each curve with energy steps of 5 eV and emission angle steps of 0.5°. The intensity histograms display the fraction of transmitted number of rays as a function of kinetic energy $E_{\mathrm{kin}}$. The maxima are normalized to 1. There may be beam clipping at the entrance aperture in GP1, depending on the aperture size and the photon footprint on the sample. The profiles are nearly symmetrical with a FWHM of ~70 eV for large apertures. However, at 0.5 mm the low energy side is steeper, and at 0.4 mm, it shows a sharp cutoff and a FWHM of ~30 eV. For small apertures, i.e. high resolution, the transmission profile shows an asymmetric shape with a steep slope at the low energy end and a longer tail at high energies.

As in any electron spectrometer, the energy bandpass depends on the pass energy $E_{pass}$ and the sizes of entrance and exit apertures. When using the instrument as a prefilter in a ToF-MM, we usually set the bandpass to 5-10 %. In the given example of $E_{pass}$ = 580 eV, the resulting bandpass width of 30 – 70 eV is sufficient to eliminate all unwanted energies from the spectrum. The final energy resolution is achieved in the ToF section. For this mode of operation a diameter of FA and SA in the range of 0.5 – 1 mm is sufficient.

By reducing the pass energy and slit size, the dodecapole setup can also be used as a dispersive energy filter without subsequent ToF analysis. The present prototype with 4° deflection angle has been experimentally tested at pass energies and aperture sizes down to $\leq$ 20 eV and 100 $\mu$m, respectively. This combination results in a resolution well below 1 eV. It is crucial for this mode that a large fraction of the photoelectrons is focused into the small circular entrance aperture. This requires a high electron flux density emitted from the sample, *i.e.* a small photon footprint. On the other hand, the spherical aberration of the lens group 1 must be minimized, in order to obtain a small image of the photon spot in the first Gaussian plane GP1. In the *k*-imaging mode, a maximum solid angle interval is aimed at, which requires a large diameter of the *k*-image in the first reciprocal image plane RP1 (see left side of Fig. 3(c)).

The primary goal of this study was to develop an easy-to-use, compact bandpass prefilter. We did not systematically vary the *deflection angle* $\alpha$ in the simulations. Nevertheless, the simulation revealed one basic property of this design: Up to moderate angles, the dispersion increases linearly with $\alpha$. Thus, the case shown in Fig. 3 but with $\alpha$ = 10° would already yield a resolution of 1 % for a slit size of 200 $\mu$m. For $E_{pass} \leq$ 50 eV this would give a resolution of $\leq$ 500 meV, which fits well with the typical bandwidths of hard X-ray beamlines. The large k-field of view is essentially retained, but the higher energy resolution at larger angles $\alpha$ is paid for by a lower *k*-resolution.

## 2.4 Implementation into a momentum microscope

The dodecapole was implemented into a high-energy momentum microscope and installed at the hard-X-ray beamline P22 of PETRA-III (DESY, Hamburg) [21]. Figure 4 shows a schematic view (a) and ray-tracing calculation (b) of the complete instrument. It comprises the front lens and 4 zoom lens groups as discussed in Figs. 1 and 3. Lens group 3 projects the *k*-image (or alternatively a Gaussian image) onto the delayline detector (DLD) via the ToF drift tube.

The optics comprises three octupoles serving as deflector/stigmator units, the first and second acting on the *k*- and Gauss-image and the third (behind the exit aperture SA of the dodecapole unit) for directing the beam parallel to the original optical axis. CA, FA and SA denote three sets of selectable and adjustable apertures, contrast aperture in the back-focal plane (reciprocal plane RP1), field aperture in the first intermediate image (in Gaussian plane GP1) and selector aperture (in plane GP2). The aperture arrays FA and SA comprise auxiliary fine TEM-grids for easy adjustment of the Gaussian planes. In the plane RP1 the k-images have diameters of up to 20 mm, hence the CA grid is a hexagonal array of small holes.

In a modified setup we placed an array of contrast apertures for PEEM mode in the conjugate plane RP2, where the size of the *k*-image is smaller. The instrument can focus real-space or momentum images of the photoelectrons on the delay-line detector (DLD), just by varying lens settings. For real-space imaging of a very large field-of-view up to >3 mm, the front lens is adjusted such that the first reciprocal image RP1 is shifted into the plane of FA. Then, RP and GP are interchanged and a real-space image is focused on the DLD. This is particularly important for ToF X-PEEM, *i.e.* spatial imaging on a core-level signal.

The front lens can be operated in different modes, as discussed in [19, 22]. In the classical *extractor mode* the first electrode is at high positive voltage. In the *zero-field mode* the extractor is 'switched off' by setting it equal to the sample potential. Finally, in the *repeller mode* the first electrode is used as a retarding lens. This mode and to some extent the zero mode suppress the main part of the space-charge interaction with the slow electrons [23].

For a total deflection angle of $\alpha$ = 4° and a distance of 120 mm between center of dodecapole and exit aperture (scaled scheme in Fig. 3(a)), the beam displacement is just 8 mm. Such a small displacement is compatible with a linear vacuum vessel and linear mumetal tube for magnetic shielding of the entire unit. Downstream of the exit aperture, a short octupole deflector / stigmator directs the beam by -4° parallel to the original optical axis. There the beam diameter is very small so that this second deflection is much less demanding than the deflection in the dodecapole, where a large *k*-image is located.

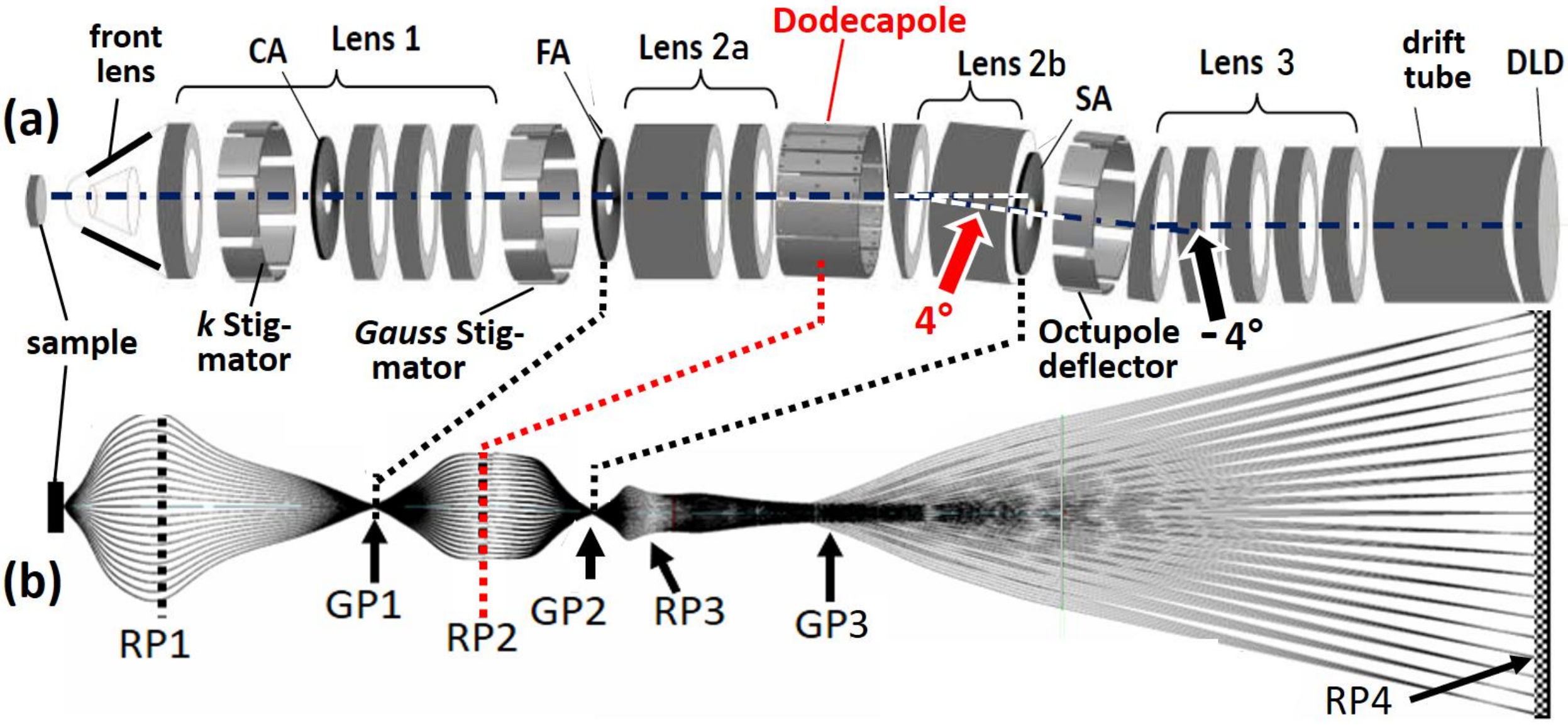

**Figure 4**
Time-of-flight momentum microscope with integrated dodecapole bandpass-filter. (a) Schematic sketch with assignment of the elements. (b) Ray-tracing calculation. CA, FA and SA denote contrast aperture, field aperture and selector aperture; the latter two are piezomotor-driven arrays of 16 apertures of different sizes. GP and RP denote the Gaussian and reciprocal image planes. The optics further includes 2 octupole stigmators (close to RP1 and GP1) for beam shaping and an octupole deflector behind GP2 directing the beam parallel to the optical axis. DLD denotes the delayline detector.

# 3. Performance of the dodecapole bandpass filter in a time-of-flight momentum microscope using hard X-ray synchrotron radiation

## 3.1 Bandwidth as function of pass energy and slit size

The performance of the setup shown in Fig. 4 was characterized at beamline P22 of PETRA III (DESY, Hamburg). The bandwidth selection was investigated at a photon energy of 5045 eV. Figure 5 shows a series of ToF spectra for different pass energies and different slit sizes. A metallic film containing vanadium and antimony was used as a test sample. The photoelectrons from the Sb $3d_{5/2}$ core level have been retarded to the pass energy $E_{pass}$ of the dodecapole, as stated in the panels. Therefore, this core level peak is the centre of the transmitted energy bands. The Sb $3d_{5/2}$ peak is framed on the low energy side by its spin-orbit partner Sb $3d_{3/2}$ at a distance of 9 eV and on the high energy side by the V $2p$ spin-orbit doublet with a distance of 16 eV between the Sb $3d_{5/2}$ and the V $2p_{3/2}$ peaks.

Figures 5(a, g) show time-of-flight spectra with peak assignments. The spectra are plotted as measured on a linear time-of-flight scale. The corresponding kinetic energy scales are shown on top of the panels, reflecting the relationship $E_{kin} \propto \tau^{-2}$ ($\tau$: time-of-flight). The entire group Sb $3d$ ($E_B$ = 528 / 537 eV) plus V $2p$ ($E_B$ = 512 / 520 eV) has a width of 25 eV, which is perfectly suited to study the confinement of the transmitted energy interval as a function of the pass energy and slit width. A number of further core levels of Sb and V at lower binding energies yield intense signals at higher kinetic energies, providing a rich scenario for observing the suppression of higher energy electrons by the bandpass filter.

The left column of Fig. 5 shows a series of time-of-flight spectra as a function of pass energy between $E_{pass}$ = 1400 eV (a) and 200 eV (f). The non-linear scale at the top shows the kinetic energy in the drift tube of the ToF section. The spectrum transmitted by the dodecapole is recorded by the subsequent ToF analysis. Here the Sb $3d_{5/2}$ peak is set to a drift energy of 60 eV. The first spectrum (a) was recorded with large slit sizes of 3 mm to visualize the high energy part of the spectrum at flight times between 0 and 50 ns, corresponding to drift energies up to 500 eV. Indeed, the group of shallow core levels of the constituent atoms appears on the left side of the spectrum. At such high drift energies, the energy scale is strongly compressed; resulting in a total spectrum of 470 eV. By reducing the slit size to 1 mm (at $E_{pass}$ = 1400 eV), the high energy component is strongly suppressed, Fig. 5(b). Stepwise reduction of the pass energy removes all electrons faster than the desired energy interval (c-f). The high-energy onset of the spectra is indicated by arrows.

The right column of Fig. 5 shows the analogous series for varying the aperture size, recorded for $E_{pass}$= 600 eV. The non-linear scale at the top shows the kinetic energy in the dodecapole. For the large aperture diameters of 4 mm / 3 mm, the spectrum extends into the region of the shallow core levels (g). In the sequence of the slit sizes 3, 2, 1.5, 0.75 and 0.5 mm (Figs. 5(h-l)), the spectrum is increasingly narrowed. The high energy onset of the spectra is marked by arrows. In the 0.5 mm spectrum (l), the onset is at 630 eV and the intensity drops to zero at about 570 eV. Assuming a shape as shown in Fig. 3(e), we estimate an effective bandpass on the order of 20 eV (FWHM), in good agreement with the simulation. As expected, the leading edge is steeper at higher energies than the trailing edge at lower energies.

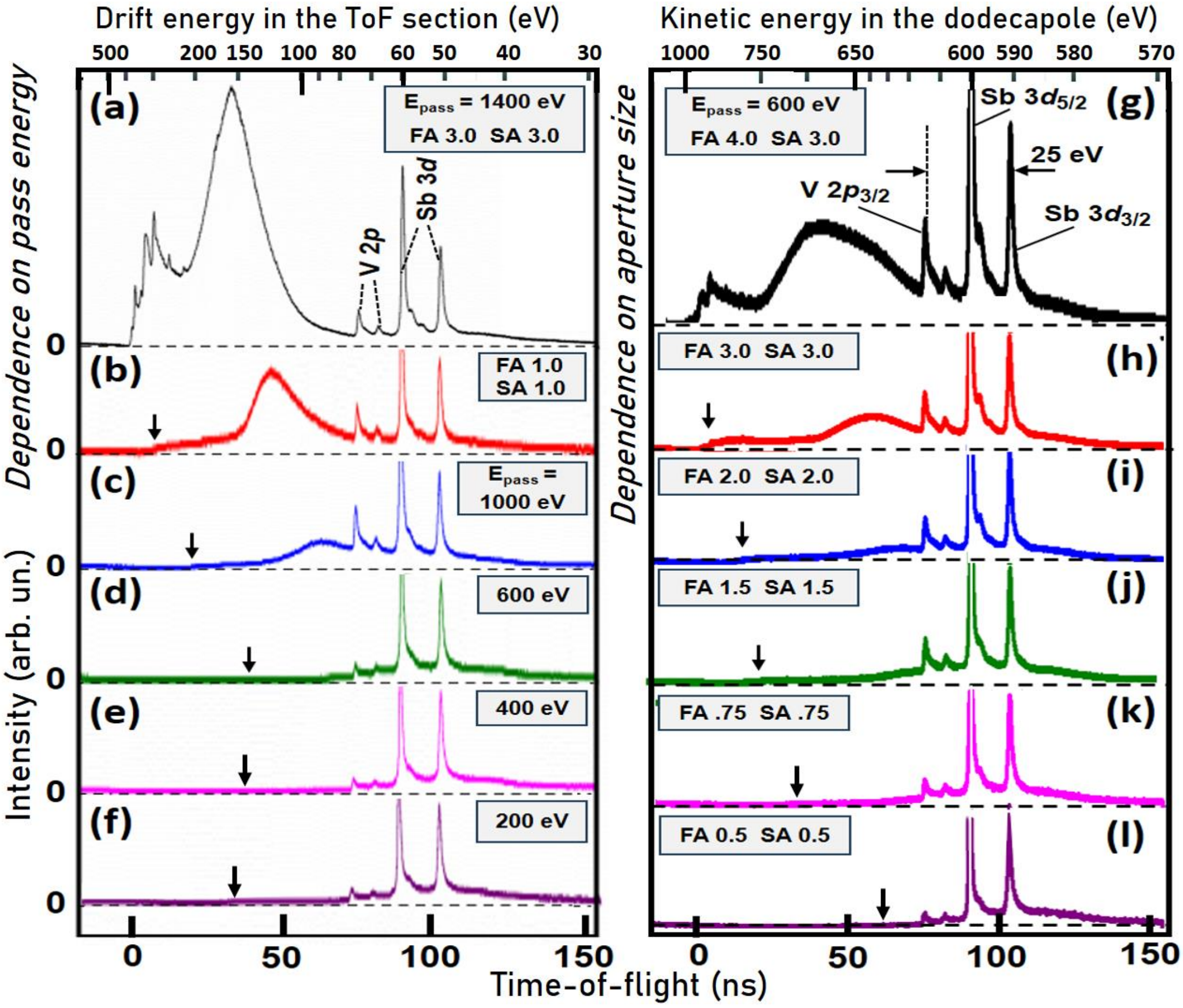

**Figure 5**

Time-of-flight spectra measured in the hard X-ray range (hν = 5045 eV) for a series of pass energies of the dodecapole bandpass filter (left column) and for different entrance and exit apertures (right column). FA and SA denote the field aperture (= entrance 'slit') and selector aperture (= exit 'slit'); the numbers 4.0 to 0.5 give the diameters in mm. In order to emphasize the smaller signals, the maximum of the main peak Sb $3d_{5/2}$ is cut in panels (b-l). The non-linear scales at the top show the drift energy in the ToF section (left column) and the kinetic energy in the dodecapole (right column).

With transmitted bandwidths between 20 and several hundred eV, the bandpass filter behaves as expected from the ray tracing simulations in Fig. 3(e). In a first attempt at even lower pass energies with a laboratory source we found bandpass widths of < 1 eV for a polycrystalline sample (without *k*-imaging).

## 3.2 Imaging properties of the dodecapole filter

It was expected that the spectral performance of the dodecapole filter would not deviate significantly from the theoretical expectation. The behavior of the electron beam in the largely homogeneous deflector field, Fig. 2(b), is reliably predicted by the simulation. Therefore, the measured values for the bandpass are in good agreement with expectations. However, it was not clear, whether the imaging performance, in particular the fringing field effects at the entrance and exit, are correctly described in the 3D model in Simion [20]. As the beam passes through the small entrance and exit apertures, the *k*-image is 'encoded' in the form of an angular pattern. Fringing fields can strongly affect this angular pattern. Extensive work has been done on to the fringing fields of hemispherical analyzers. In a classic work Jost [24] introduced a special electrode geometry and recently Tusche et al. [6] discussed the fringing field effect of the 'Jost electrodes' for hemisphere-based momentum microscopes. Due to the small deflection angle $\alpha$, our design of the dodecapole filter does not include electrodes for fringing field correction. However, until the first experiments, it remained unclear whether the dodecapole would lead to significant image aberrations.

We demonstrate the imaging performance of the bandpass filter for two key applications of the high-energy momentum microscope at beamline P22: (i) full-field *core-level photoelectron diffraction* and (ii) *bulk valence band mapping*. Figure 6 shows a selection of core-level diffractograms for Si 1*s*, 2*s* and 2*p* (a, b, c) and Ge 2*p* and 3*p* (e, f, g) with the corresponding ToF spectra (d, h). The pass energy of the dodecapole was 400 eV. At these energies the XPD patterns exhibit pronounced Kikuchi diagrams. These are rich in detail and show superior contrast and resolution compared to previous measurements on Si [25, 26], Ge [27] and GaAs [28]. We attribute this partly to the blocking of electron paths far from the optical axis combined with a 20-30% smaller *k*-field of view and partly to the background reduction provided by the dodecapole prefilter. The latter effect is also visible in the spectra, which show virtually no background.

In these examples the countrate in the DLD was >$10^6$ counts per second. The photoelectron diffraction patterns could then be observed in real time at a frame rate of a few seconds. Without the bandpass filter, the background is much higher and the majority of electrons reaching the DLD do not come from the core-level of interest. In turn, the saturation of the DLD limits the recording speed. One of the most important applications of full-field XPD using a momentum microscope is the structural and spectroscopic analysis of dopants in semiconductors; see examples of Mn in GaAs [28] and Te in Si [26]. In these cases, the core-level signals of the dopants are quite weak, so the background suppression is even more important. However, the main advantage of the bandpass is the elimination of higher order artifacts such as those shown in Fig. 1(e,f). Note that the final energy resolution is not affected by the dodecapole filter as it is defined by the time resolution and the drift energy in the ToF analyzer.

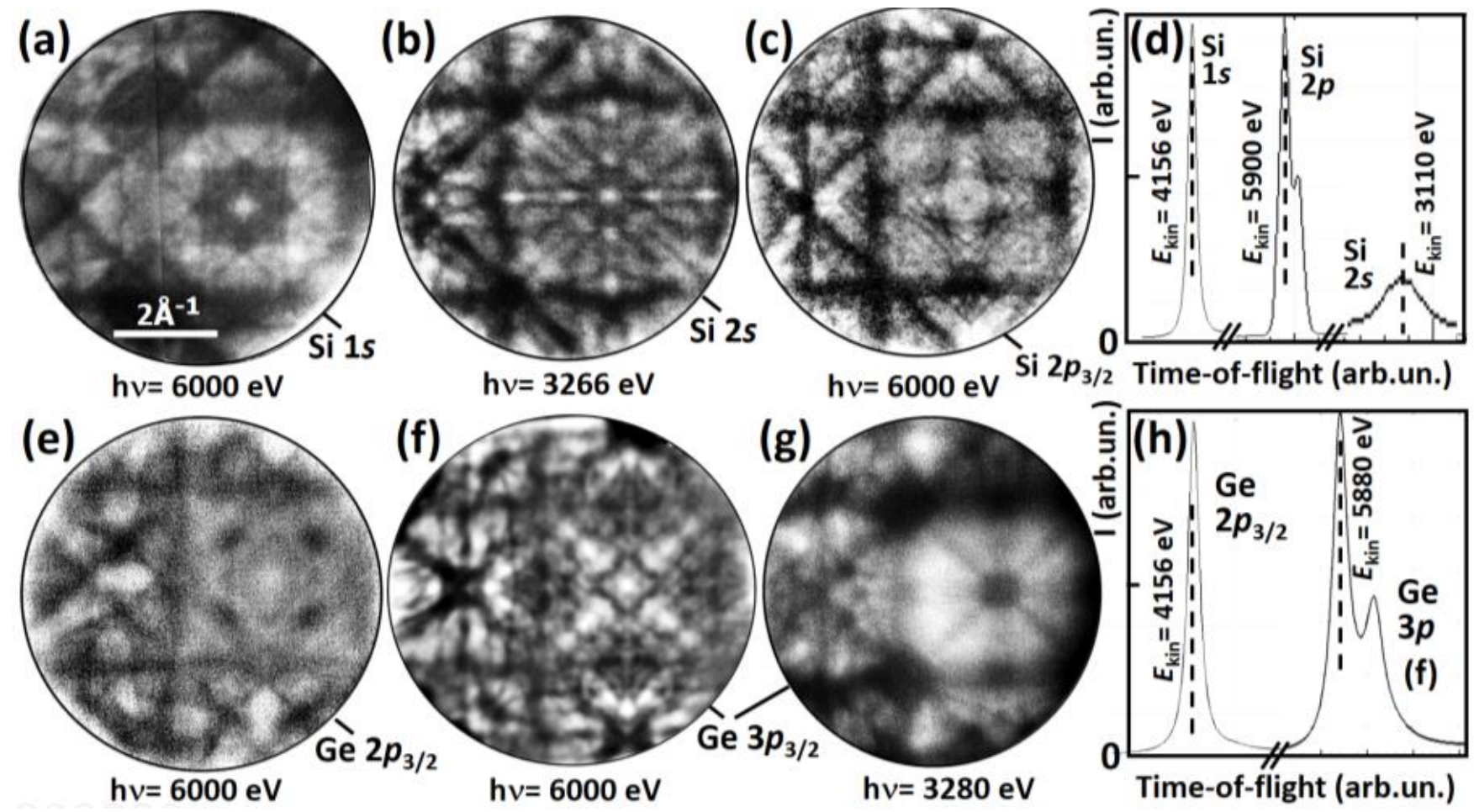

**Figure 6**

Hard X-ray core-level photoelectron diffraction patterns measured at beamline P22 of PETRA III. Top row silicon 1*s*, 2*s* and 2*p* and bottom row germanium 2*p* and 3*p* patterns, recorded at hν = 6 keV (a, c, e, f), 3.266 keV (b) and 3.28 keV (g). (d, h) Corresponding time-of-flight spectra with corresponding energies $E_{kin}$.

As an example of bulk valence band (VB) mapping, we have chosen silicon. Despite it's high Debye temperature, $\Theta_D \approx 680$ K, this light element (Z = 14) should not be suitable for hard X-ray ARPES (HARPES). The Debye-Waller factor is considered to determine the ratio of indirect to direct transitions. For Si this ratio already reaches 0.5 at h$\nu \approx 1$ keV and 20 K, see Fig. 30 in ref. [29]. This regime is called the 'XPS limit' [29] because phonon scattering processes randomize the *k*-information while almost preserving the energy, i.e. the XPS spectra.

Figure 7 shows a sequence of $k_x$-$k_y$ sections for Si(001) recorded at h$\nu$ = 3420 eV (a-g) and 3266 eV (h-l). These energies were chosen because in 3D *k*-space the photoemission final state sphere in the 13$^{th}$ repeated BZ intersects the high-symmetry plane $\Gamma$KX (at 3420 eV) and the plane crossing the L-points, halfway between $\Gamma$ and the rim of the BZ (at 3266 eV). Here we only demonstrate the quality of the band structure features and discuss some peculiarities of hard X-ray photoemission of light elements; a more detailed analysis is in preparation [30]. The first row shows the VB patterns at binding energies $E_B$ between 0 and 5.4 eV (with respect to the VB maximum). The series shows the topmost band visible as a single dot in (a), opening to an outwardly dispersing band (fourfold symmetry) with increasing $E_B$ (b-f). The dispersions along the $\Sigma$ and $\Delta$ directions are shown in the $E_B$-*vs*-$k_\parallel$ plots, Figs. 7(o, p). The uppermost band looks sharp up to its top, which defines the VB maximum. The $k_x$-$k_y$ patterns at the same $E_B$ values in the second row (h-l) look very different, because the different photon energy leads to another plane in the 3D BZ, that cuts through the L points. The corresponding $E_B$-*vs*-$k_\parallel$ section in Fig. 7(q) shows that in this plane the bands do not reach the VB maximum. In Fig. 7(p) the split-off band with its minimum at the X-point is clearly visible, all in agreement with calculations for Si [31]. In agreement with Strocov et al. [32] we do not find any traces of multiband final states for Si.

It is noteworthy that all images show not only the bright dispersing band features, but also a highly structured background consisting of a symmetrical pattern of distinct dark lines that show no dispersion; cf. Figs. 7(a-g) and (h-l). This background is caused by quasi-elastic phonon scattering processes and has been observed and discussed in previous HARPES experiments. It has a characteristic spectral shape reflecting the *matrix element weighted DOS* (MEWDOS) [29, 33, 34]. The MEWDOS background is visible in the form of diffuse bright regions in the $E_B$-*vs*-$k_\parallel$ plots, Figs. 7(o, p, q). In electron microscopy the origin of this background is called 'thermal diffuse scattering' and has been studied in detail ([35] and references therein). The series in Figs. 7(a-l) confirm our earlier observation for several elements and compounds [36] that this background of indirect transitions is by no means 'diffuse', but shows a pronounced structure. A comparison of (e) and (g), both for $E_B$ = 4.4 eV but sample temperatures 30 and 190 K, respectively, reveals that this background increases with increasing temperature, at the expense of the direct transitions. The electrons that 'drop out' of the coherent final state wave field due to a scattering process experience Kikuchi and Laue diffraction as they escape from the crystal; a detailed discussion is given in [37].

Close inspection of the imprinted Kikuchi patterns reveals that the structured background in Figs. 7(c-e) and (h-k) is significantly different, reflecting the different wavelength of the electrons at the two energies. A qualitative comparison of Figs. 7 and 6 also shows that the richness in detail of the VB and core-level Kikuchi patterns is very similar. The VB patterns after multiplicative correction by the Kikuchi pattern are shown in Fig. 7(m, n); for details of this procedure, see [36]. Finally, we note that the modulation by the Kikuchi pattern is even visible in the band features themselves, as marked by arrows in Figs. 7(e, f, g, i, l). Fine dark Kikuchi lines modulate the VB ring in (e, f, i, l) and the light-dark transition at the edge of the primary rectangular Kikuchi band suppresses the intensity of the VB ring in (g). In agreement with DOS calculations [38], the MEWDOS background shows a deep minimum between $E_B$ = 5 and 6 eV, visible as a dark horizontal stripe in Figs. 7 (o, p, q). This has a very positive effect on the $k_x$-$k_y$ patterns at $E_B$ = 5.4 eV, where the Kikuchi background is almost invisible. The band patterns show excellent contrast even without correction and their modulation by sharp dark Kikuchi lines is striking, arrows in Figs. 7(f, l).

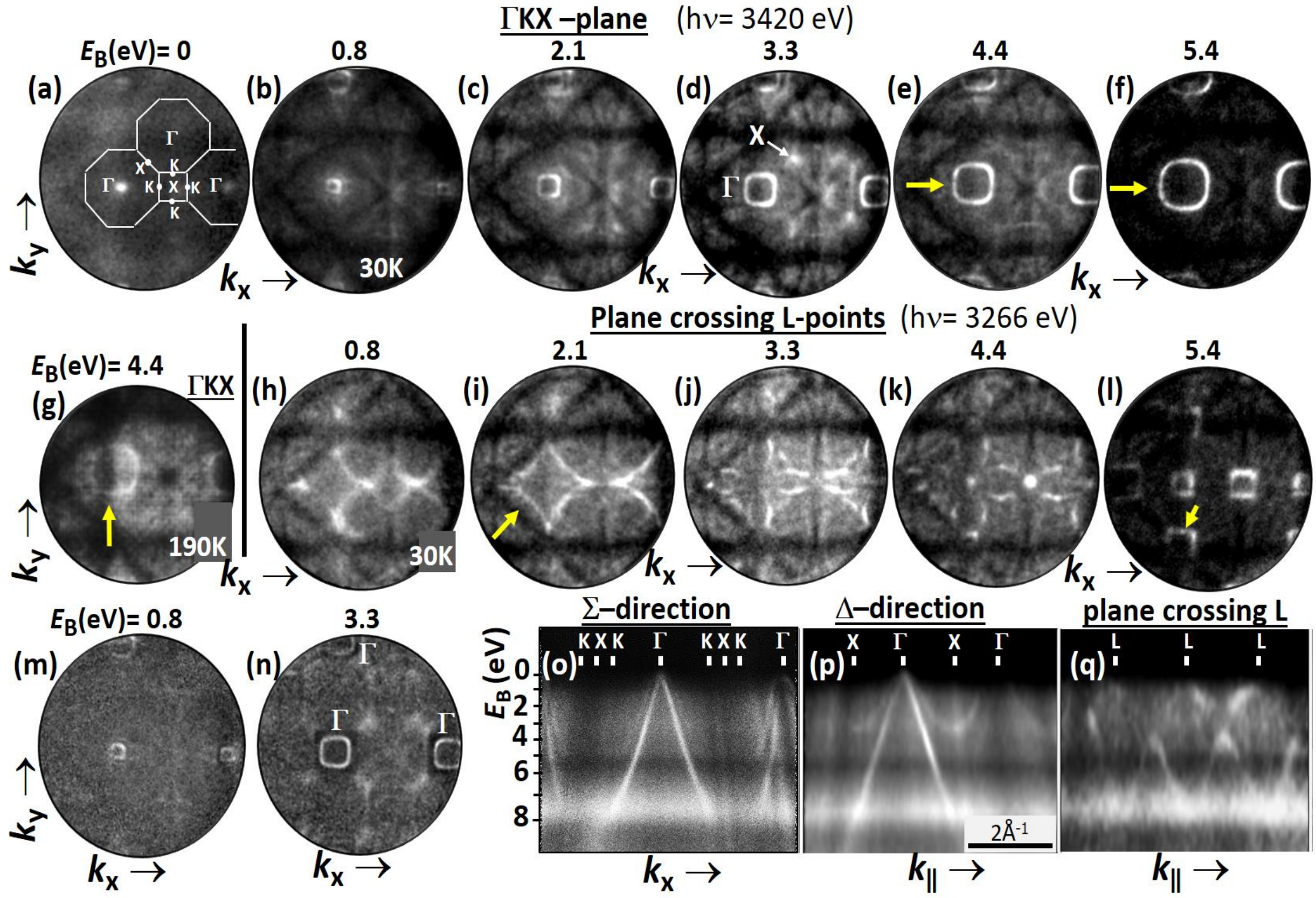

**Figure 7**

Hard X-ray valence-band patterns of Si(100), recorded at beamline P22 of PETRA III using a pass energy of the dodecapole of $E_{pass}$ = 600 eV (sample temperature 30 K except (g)). (a-f) Series of $k_x$-$k_y$ patterns; $E_B$-values are given on top of the panels. $E_B$ is referenced to the valence-band maximum. At hν = 3420 eV (top row) the direct transition leads to the ΓKX-plane of the 13$^{th}$ repeated Brillouin zone (BZ) along $k_z$. (g) Same as (e) but for 190 K. (h-l) Same binding energies as (b-f) but at hν = 3266 eV, leading to the plane intersecting the L-points of the 3D BZ. (m, n) Patterns (b, d) after eliminating the imprinted Kikuchi diffraction by multiplicative correction. (o-q) $E_B$-*vs*-$k_\parallel$ cuts along the Σ and Δ direction and in the plane through the L points. Except (m, n), all panels show raw data as measured.

## 4. Summary and Conclusion

In time-of-flight instruments, photoelectrons with higher energy than the electrons of interest can pass through the spectrometer and reach the detector, similar to a high-pass filter. If the time-of-flight τ of the faster electrons differs by an integer multiple of the period *T* of the photon pulses, these electrons appear in the same time window as the true signal. The detector cannot distinguish electrons with τ differing by ± *nT* (*n* natural number). This phenomenon is commonly referred to as 'temporal aliasing'. In momentum microscopy (MM) artifact signals from this effect can be superimposed on the true signal [17]. Temporal aliasing is particularly severe in the X-ray region, where the energy spectrum is very broad and higher order contributions from the monochromator and undulator release electrons with even higher energies. In addition to creating artifacts, such unwanted electrons can also lead to detector saturation.

Here we described a compact bandpass prefilter that eliminates electrons with energies above or below the desired range before the beam enters the time-of-flight analyzer. The heart of the device is an electrostatic dodecapole with an asymmetric electrode arrangement that allows energy-dispersive beam deflection and correction of image aberrations up to 3$^{rd}$ order. The electrode configuration is designed so that the energy-dispersive dipole field is homogeneous over a large part (~80%) of the cross-sectional area. The filter is framed by transfer lenses in the input and output branches. Thanks to a beam deflection angle of only 4°, the entire instrument is very compact and housed in a straight mumetal vacuum tube. This type of background reduction also works with other types of analyzers [39]. The filter can be integrated into MMs or PEEMs for imaging in real or reciprocal space.

We presented results for a dodecapole filter in a high energy MM end station at the hard X-ray beamline P22 at PETRA III (DESY, Hamburg). Systematic measurements for pass energies between 200 and 1400 eV and entrance and exit apertures between 0.5 and 4 mm (on two piezomotor driven arrays with 16 apertures each) confirmed the expected spectral confinement and recorded transmitted energy intervals between 20 eV and several hundred eV. In this *preselector* mode, artifacts caused by higher energy electrons as well as slow secondary electrons are effectively eliminated. In this beamline, the photon footprint is small, allowing a large fraction of the electrons within the desired energy band to pass through the entrance and exit apertures. These are the same criteria that have been discussed in the context of a hemispherical analyzer as a prefilter [7, 8].

In conclusion, an electrostatic dodecapole framed by two transfer lenses and arrays of entrance and exit apertures can serve as a simple but effective bandpass filter. For the use as a prefilter in a ToF spectrometer, a deflection angle of 4° proved to be a good compromise, providing sufficient resolution, negligible aberrations and a small beam shift of few mm (compatible with a straight vacuum housing). First tests with a low-energy laboratory source showed a bandpass of < 1 eV, allowing spectroscopy without subsequent ToF analysis. Here we have shown examples where the final energy resolution is defined by the ToF analysis. The electron-optical system shown in Fig. 4 enables *momentum microscopy*, *full-field photoelectron diffraction* and *high resolution X-PEEM* (first results in [19]). *Sub-micron HARPES* exploits the confinement of the probe spot by small field apertures. The ultimate performance limit is posed by the space-charge interaction, which can be reduced by a novel front lens architecture. The front lens of the instrument combines the conventional *extractor mode* (for maximum *k*-field of view) with the *zero-field mode* (for 3D structured samples) and the *repeller mode* (for space charge suppression) [19, 22]. The dodecapole improves the signal-to-background ratio and eliminates signals from higher order admixtures in the photon beam. The instrument operates for kinetic energies up to $E_{kin}$ > 8 keV.

## Acknowledgements

We acknowledge DESY (Hamburg, Germany), a member of the Helmholtz Association HGF, for the provision of experimental facilities. Parts of this research were carried out at PETRA III, beamline P22. Special thanks go to Surface Concept GmbH, Mainz, for technical support with the delay line detector. The development was funded by BMBF (Projects 05K19UM1 and 05K22UM2) and Deutsche Forschungsgemeinschaft DFG (German Research Foundation) through TRR 173-268565370 Spin +X (project A02), Project No. Scho341/16-1 and TRR 288-422213477 Elasto-Q-Mat (project B04).

## Data availability

All data shown within this article is available on reasonable request. We declare no conflict of interest.